\documentclass{article}
\usepackage{arxiv}
\usepackage[utf8]{inputenc} 
\usepackage[T1]{fontenc}    
\usepackage{hyperref}       
\usepackage{url}            
\usepackage{booktabs}       
\usepackage{amsfonts}       
\usepackage{nicefrac}       
\usepackage{microtype}      
\usepackage{apacite}
\usepackage{multicol}
\usepackage{multirow}
\usepackage{lipsum}
\usepackage{graphicx}
\DeclareUnicodeCharacter{2060}{\nolinebreak}

\title{The Consistency of Trust-Sales Relationship in Latin American E-commerce}

\author{
Juan C. Correa\\
CESA Business School\\
Bogotá, Colombia\\
\texttt{juan.correan@cesa.edu.co} \\
\And
Henry Laverde-Rojas \\
Fundación Universitaria Konrad Lorenz\\
Bogotá, Colombia\\
\And
Camilo A. Martínez \\
Fundación Universitaria Konrad Lorenz\\
Bogotá, Colombia\\
\And
Oscar J. Camargo \\
Fundación Universitaria Konrad Lorenz\\
Bogotá, Colombia\\
\And
Gustavo Rojas-Matute\\
American University\\
Washington, USA\\
\And
Marithza Sandoval-Escobar\\
Fundación Universitaria Konrad Lorenz\\
Bogotá, Colombia
}

\begin{document}
\maketitle
\begin{abstract}
Customer's trust in vendors' reputation is a key factor that facilitates economic transactions in e-commerce platforms. Although the trust-sales relationship is assumed robust and consistent, its empirical evidence remains neglected for Latin American countries. This work aims to provide a data-driven comprehensive framework for extracting valuable knowledge from public data available in the leading Latin American e-commerce platform with commercial operations in 18 countries. Only Argentina, Brasil, Chile, Colombia, Ecuador, Mexico, Uruguay, and Venezuela showed the highest trust indexes among all nations analyzed. The trust-sales relationship was statistically inconsistent across nations but worked as the most important predictor of sales, followed by purchase intention and price.
\end{abstract}


\section{Introduction}
Trust is probably one of the most systematically researched terms in the literature of electronic commerce. According to \citeauthor{Hsu2015} \citeyear{Hsu2015}, the United States of America, China, Taiwan, or the UK dominated the research on trust in e-commerce between 2000 and 2014. From 2015, there exist studies of trust in almost all countries, of which South Korea \cite{Joo2015,Jeon2021}, Indonesia \cite{Sfenrianto2018}, India \cite{Kaushik20201,Aeron201945}, Pakistan \cite{Butt2021}, Qatar \cite{Al-Khalaf2020}, Lebanon \cite{Dabbous2020}, or USA \cite{Furner2021} emerged as a few recent examples. 

From a geographic point of view, the evidence provided by \citeauthor{Hsu2015} \citeyear{Hsu2015} seems to pinpoint that the research on customers' trust in e-commerce sellers from Latin America is far from being systematic and well documented despite the several references available from this region \cite{changchit2009,Bianchi2012,dakduk2017,Sanchez2017,LopezJimenez2021}. Above and beyond this circumstance, the specific analysis of the trust-sales relationship remains neglected in the literature. 

The current study aims to argue a data-driven framework that harnesses the advantages of web scraping procedures \cite{correa2019,Teichert2020}. Such procedures are powerful in terms of extracting valuable knowledge from public data available in Mercadolibre, the leading Latin American e-commerce platform with commercial operations in 18 countries \cite{hortaccsu2009,Benedict2021}. In the remainder of this article, the organization is as follows. The following section has two parts. The first part summarizes existing theoretical approaches of trust in e-commerce. The second part briefly summarizes the Latin American context to introduce the empirical elements on which the data-driven framework is built upon. The core of this framework relies on the operationalization of trust calculated by the platform. It will be immediately evident that this index overcomes many conceptual and operational confusions related to trust highlighted by previous studies \cite{Grabner2003,Connolly2008}. This index aims to quantify the interaction that each seller establishes with each customer in the platform. While the purpose of this index seems justified in terms of the sensitive information it conveys for future customers, other metrics from the platform are also available but remain unstudied as we are not aware of previous efforts done in the academic community. This paper leverages these other metrics to estimate its predictive role for sales. 

The empirical elements analyzed will show a straightforward connection with the section of materials and method. The contribution provided here should be evident in at least two orientations. First, it provides an alternative methodological approach that deviates from the dominant survey-based instrumentation deeply rooted in psychometrics of trust in e-commerce. This methodological treatment is, in our perspective, consistent with a business data analytic framework \cite{Provost2013} that we regard as a powerful complement to more traditional frameworks about trust. Second, this paper is among the first ones in providing fresh evidence that shows how consistent trust is across Latin American nations. The statistical techniques employed for analytical purposes are provided as supplemental material so others can use it by following the standards of reproducible research \cite{Gandrud2018}.

\section{Literature Review}
Trust is an essential factor for customers decision on purchasing from sellers on the Internet \cite{Hoffman1999,Cheung2000,Corbitt2003}. Despite the simplicity of this idea, the scientific study of customers' trust has called for interdisciplinary efforts, including human-computer studies, business and management, psychology, neuroscience, economics, sociology, anthropology, political sciences and information sciences \cite{Grabner2003,Connolly2007,Connolly2008}. Nowadays, the shared interest between these disciplines has become cyberbehavior as the object of study related to the effects of the Internet and cyberspace on individuals and groups \cite{Serafin2019}. 

Customers' trust is observable at different scales, often applied in different practical scenarios. Some examples are useful for illustrating this statement. At the neurophysiological scale, \citeauthor{Hubert2019} \citeyear{Hubert2019} analyzed how customers' trust in the textual descriptions of products sold in eBay relates to a particular type of brain activity that revealed significant differences between impulsive and non-impulsive buyers. The neurological structures implied in these differences were the dorsolateral, prefrontal cortex, the island region and the caudate, which belong to the striatal area, also known as the ``reward center.'' This last structure, in particular, was recruited significantly more from non-impulsive buyers than from impulsive buyers. At the self-reported behavioral scale, \citeauthor{Ramadan2019} \citeyear{Ramadan2019} analyzed the effects that Amazon Prime Now had on shoppers' impulsiveness and addiction and reported that clients' gratification was due to its customer service, which in turn facilitated an established trust and love with the retailer. At a nationwide scale, \citeauthor{Qu2015} \citeyear{Qu2015} analyzed the impacts of trust on open and closed business-to-bussiness (B2B) e-commerce. The analysis from 27 European and related countries revealed that social trust in a nation promoted the use of open B2B e-commerce, while impeding the use of closed B2B e-commerce. Besides, social trust negatively moderated the relationship between ﬁrm-level IT experience and closed B2B e-commerce. 

As the previous references show, the empirical analysis of customers' trust depends on the following four criteria:  1) the disciplinary perspective (e.g., psychology, neuroscience, economics), 2) the scale of observation (e.g., neurophysiological, behavioral, transactional, self-reports), 3) the measurement instrumentation employed (e.g., surveys, standardized questionnaires, functional magnetic resonance imaging techniques), and 4) the contextual scenario of the study (e.g., retailers' websites, e-commerce platforms, digital apps). These criteria might be regarded as the building blocks of a typology \cite{McKnight2001}. According to \citeauthor{Jaakkola2020} \citeyear{Jaakkola2020}, a typology often starts with identifying an important but fragmented research topic that demands a meaningful knowledge organization that helps one understand where a particular study fits in. Although this typological understanding of trust has its own pedagogical value, it deserves the exposition of the ideas derived from existing theoretical approaches. The following section aims to summarize these theoretical foundations. 

\subsection{Theoretical Approaches of Trust in e-commerce}

As per \citeauthor{Grabner2003} \citeyear{Grabner2003}, two well-known and widely accepted theoretical frameworks are relevant in the literature of trust. On the one hand, the theory of planned behavior (TPB) by \citeauthor{ajzen1991} \citeyear{ajzen1991} posits that individuals intentions to perform behaviors can be predicted with accuracy from their attitudes toward their behavior, the subjective norms shared by the members of their social groups, and their perceived behavioral control. These intentions, in turn, explain a considerable amount of variance in actual behavior. On the other hand, the technology acceptance model (TAM) by \citeauthor{davis1989} \citeyear{davis1989} states the existence of two psychological factors that influence the acceptance of technology innovations. These factors are “perceived usefulness” and “perceived ease of use”. The former refers to the perception of a user about the subjective probability that the use of technology will help increase his performance. The latter refers to the individual’s subjective appreciation that using a particular technology involves least efforts. Another related model, widely disseminated, is the so-called Unified Theory of Acceptance and Use of Technology (UTAUT), proposed by \citeauthor{Venkatesh2003} \citeyear{Venkatesh2003}. This last model posits four dimensions (i.e., performance expectancy, effort expectancy, social influence, and facilitating conditions) and four moderators (i.e., age, gender, experience, and voluntariness). It is said that these dimensions and moderators play a major role in predicting the intention to use a particular technology \cite{Venkatesh2016}. Regardless of their conceptual differences, UTAUT, TPB, and TAM have gained considerable evidence that supports them as relevant ones to understand the complexities of the technology-mediated relationships between buyers and sellers through e-commerce platforms \cite{Alkhunaizan2012,Musleh2015}. In fact, according to \citeauthor{correa2019} \citeyear{correa2019}, the integration of these perspectives became a well-accepted proposition since \citeauthor{pavlou2003} \citeyear{pavlou2003} and \citeauthor{gefen2003} \citeyear{gefen2003} posited the role of trust and risk as the psychological mechanisms that make this theoretical integration a pertinent one for the study of e-commerce. 

In Pavlou's terms, the primary constructs for capturing consumer acceptance of e-commerce are the intention to transact and the on-line transaction behavior. These constructs, however, are related to trust \cite{suh2003,McKnight2001} and perceived risk \cite{Glover2010} because of the implicit uncertainty in the e-commerce environment. Pavlou pointed out two forms of uncertainty in on-line transactions. The first one is called ``behavioral uncertainty'' related to the actions of the web retailers and the second is known as ``environmental uncertainty'' associated with the unpredictable nature of the Internet. The perceived associated risks to these forms of uncertainty posit some challenges for current research on e-commerce, as we will show immediately. Table \ref{T1}, summarizes these risks and illustrates them with some examples that are admittedly compatible with other frameworks \cite{Cheung2000,Shankar2002,Connolly2007,Connolly2008}.

\begin{table}[h!]
\centering
\scriptsize{}
\caption{Uncertainties and associated risks in e-commerce transactions} 
\label{T1}
\begin{tabular}{cll}
\hline
\multicolumn{1}{l}{\begin{tabular}[c]{@{}l@{}}Type of \\ uncertainty\end{tabular}} & \begin{tabular}[c]{@{}l@{}}Associated \\ risk\end{tabular} & Example \\ \hline
\multirow{4}{*}{Behavioral} & Economic & \begin{tabular}[c]{@{}l@{}}A monetary loss that takes place when \\ a buyer pays for a product or service \\ without receiving it from the seller.\end{tabular}\\
& Personal & \begin{tabular}[c]{@{}l@{}}When sellers offer unsafe or untested\\ products that could create a buyer’s\\ health or individal issues.\end{tabular}\\
& Seller performance & \begin{tabular}[c]{@{}l@{}}Imperfect or unstandardized processes\\ of monitoring and delivering that\\ prevent consumers’ total satisfaction.\end{tabular}\\& Privacy & \begin{tabular}[c]{@{}l@{}}Opportunity to disclose personal \\ consumer information, such as\\ credit card number or address.\end{tabular} \\ \hline
\multirow{2}{*}{Environmental}                                                     & Economic & \begin{tabular}[c]{@{}l@{}}A monetary loss that takes place\\ when the transfer was not \\ successfully received by the \\ seller although it was \\ successfully discounted from\\  the buyer’s bank account.\end{tabular} \\
& Privacy & \begin{tabular}[c]{@{}l@{}}A possibility of theft of private\\ information or illegal disclosure \\ (e.g., e-mail or E-banking \\ passwords).\end{tabular}                                                        \\ \hline
\end{tabular}
\end{table}

A common practice to lessen the uncertainties in e-commerce platforms is the use of the so-called ``reputation mechanisms'' \cite{macinnes2005,Josang2007,utz2009}. Such mechanisms are effective deterrents of disputes or undesirable behaviors from both buyers and sellers in online commerce platforms. On eBay, for example, a dispute occurs when a transaction generates a buyer or seller dissatisfaction as expressed through neutral or negative feedback. 

The so-called ``electronic word-of-mouth'' (ewom) is another reputation mechanism as it encompasses reviews, ratings, and recommendations which are common in the well-known ``social commerce'' \cite{huang2013,hajli2017} where people are allowed to post their opinions about the product and services they transact in online channels. According to \citeauthor{bag2019} \citeyear{bag2019}, ewom is one of the three main online information sources that influence online shopping (being the other two the information of manufacturer and retailers and the information provided by neutral/third parties). As per \citeauthor{amblee2011} \citeyear{amblee2011}, ewom should be considered as the first and perhaps primary source of social buying experience because of its effects on consumers attitudes. Following this reasoning, \citeauthor{pang2016} \citeyear{pang2016} proposed the concept of ``review chunking'' (i.e., grouping reviews by valence), as a newly recognized factor that influences the persuasive effect of online reviews, and examined the differential effects of review chunking on product attitude for consumers with high versus low motivation to think. Their results showed that for consumers with low motivation to think, review chunking has a negative effect on product attitude, while the effect of review chunking on consumers with high motivation to think, depends on whether they read positive or negative reviews first.

The theoretical relevance of reviews, ratings, and recommendations, relies on two interrelated aspects. They have shown not only significant effects on the attitudes of consumers \cite{zloteanu2018}, but they also work as predictors of sales in e-commerce platforms \cite{cui2012,liang2015, pang2016}. For example, \citeauthor{cui2012} \citeyear{cui2012} observed 332 new products from Amazon.com over nine months and concluded that the valence of reviews and the volume of page views had a stronger effect on search products. However, the volume of reviews was more important for experience products although it had a significant impact on new product sales in the early period with a decreasing effect over time. Moreover, the percentage of negative reviews had a greater effect than that of positive reviews. \citeauthor{liang2015} \citeyear{liang2015} examined the effect of textual consumer reviews on the sales of mobile apps, and found that although consumers’ opinions on product quality occupied a larger portion of consumer reviews, their comments on service quality had a stronger unit effect on sales rankings.

The aforementioned approaches, briefly summarized, highlight the complex nature of customers' trust in vendors' reputations. Part of this complexity relates to how relevant data at a particular observation scale ends up being intertwined with relevant data at a different scale \cite{Correa2020}. Apart from these ideas, the so-call ``sharing economy'' has renewed the relevance of the relationship between trust, purchase intention, and sales \cite{Furner2021}. Roughly speaking, the ``sharing economy'' is an emergent global trend consisting of individuals' interest in access to rather than owning products or services. According to \citeauthor{correa2019} \citeyear{correa2019}, the sharing economy is intrinsically related to the concept of collaborative consumption that takes place when people coordinate the acquisition and distribution of a resource which is observable in rating, reviews, and recommendations provided by customers who acquired a product and decided to help other consumers by posting their experiences with it. From this perspective, it is evident that customers' trust is influenced by this type of information, and this influence has an impact on sales as well \cite{Chevalier2006}. Based on this idea, several approaches have evolved as alternatives to traditional frameworks that rely on surveys or questionnaires. These alternative approaches have been categorized as ``data-driven'' because they rely on the data (also known as contents) that online platform users generate \cite{Teichert2020,Suganya2020,Correa2020}. The following section aims to describe a data driven approach for customers' trust in Latin America.

\subsection{Customers' Trust in Latin America: A data-driven approach}

Currently, Latin America presents several characteristics that make it unique as an emerging economy. As per \citeauthor{Gouvea2021} \citeyear{Gouvea2021}, this region has 350 million mobile internet subscribers and 260 million more could be included as they live in a networked environment, but are not using mobile internet. Several e-commerce platforms are now available in the region, but Mercadolibre stands as the leading e-commerce website with commercial operations in 18 countries \cite{Benedict2021}⁠.

Compared with other e-commerce platforms, a distinctive feature of Mercadolibre is that customers and vendors use the local currency of their nations for conducting economic transactions. From a business data analytics perspective \cite{Provost2013}, this feature allows the analyst to grasp the size of each local market, as captured by the number of transactions for a given item in each country. As vendors in this website have to create a local account to conduct their transactions in each specific country, these accounts are geographically differentiated from the first time the user creates an account. Since Mercadolibre allows the written interaction between its users, customers can post questions to sellers regarding product presentations and warranties, shipment conditions, etc. The buyers-sellers communication entails the questions that customers ask vendors before initializing a  transaction and the answers that sellers provide in return. Once any customer decides to purchase a product sold by a seller and the platform verifies the successful transfer into the seller’s bank account, this transaction adds up into the seller’s statistics. This transaction, in turn, affects the seller's reputation as the platform encourages the customer to evaluate the seller's conduct as either positive feedback, neutral feedback, or negative feedback. The index of trust quantified by Mercadolibre takes into account this feedback.

According to the official information of the platform, Mercadolibre's trust index captures the position that a particular seller occupies in the rank of all sellers. Nonetheless, the calculation of this index also takes into account the following criteria. First, it includes the number of sales with customers' complaints. Second, it includes the shipment time. Third, it includes the number of canceled sales by the vendor. Fourth, the index updates daily by counting all conducted sales during the last 60 days (for newcomers with less than 4 months of commercial history) or the last 365 days (for experimented sellers with more than four months of trading). These criteria are used to classify sellers in the following three groups. ``Green sellers'' have sales with a maximum of 4\% of customers' complaints, 15\% of delayed shipments, and 3\% of canceled sales. ``Yellow sellers'' have a maximum of 6\% of customers' complaints, 20\% of delayed shipments, and 7\% of canceled sales. Finally, ``Orange sellers'' have a maximum of 10\% of canceled sales, 30\% of delayed shipments, and 8\% of sales with customers' complaints. Given the rules for this index calculation, its numeric expression ranges between 0 and 100\% with a negative skewness (i.e., the majority of sellers have trust index in the range between 75 and 100\%). Sellers with a systematic index below this range receive a warning notification and, in case of continuing with poor service, are suspended so they can sell no more. Another interesting characteristic that is not explicitly considered in the calculation but is highly appreciated by customers is the seller's communication in answering customer's questions about the product. In this regard, the best vendors provide quick and straightforward answers to any customer's question. This interaction is significant because it reflects the purchase interest from the customer and a pre-sale service from the vendor.

From a theoretical viewpoint, the index presents some features that make it unique. Its operationalization is remarkably different from the typical psychometric-inspired measurement that calculates a total score from the sum of points associated with each item's answer to a survey or a questionnaire. The index overcomes many conceptual and operational confusions summarized in the literature \cite{Connolly2008}. Its purpose is to reflect what exactly occurred with each particular sale. By defining an ``ideal standard'' that clearly reflects the seller's performance with flawless sales (i.e., sales of products shipped on time without malfunctioning issues and customers' complaints), it sanctions seller's deviations from this standard. A critical reader might say that this index is not consistent with the concept of customer's trust as described by the set of characteristics that \citeauthor{Connolly2007} \citeyear{Connolly2007} summarized in the past (e.g., competence, integrity, benevolence, openness and honesty). Nonetheless, the index aims to promote a generalized trust on the platform by providing an objective metric of each seller. This metric is displayed by the platform and any customer can read and understand its meaning. Another criticism is that this index has no theoretical value because it was developed regardless of existent theoretical approaches heavily instrumented in psychometric-like procedures. In our opinion, such a heuristic development is more objective and can be taken as a resource for the criterion validity of tests aiming to measure customers' trust \cite{Cronbach1955}.

\section{Materials and method}
Our methodological framework exploits the benefits of web mining \cite{correa2019} as a convenient means for the automatic extraction of relevant data for e-commerce research. We developed a computational script in the R environment \cite{RCore2019} with the aid of the ``rvest'' package \cite{Wickham2019} which was designed as a tool to extract data from web pages and deploy it as tidy data frames that are ready to analyze from a statistical point of view.

\subsection{Data Source}
As we previously mentioned, our primary data source was Mercadolibre.com. Because users' information is private according to the website, we did not conduct any procedure that vulnerates their privacy and confidential data (e.g., addresses, credit cards numbers, e-mail accounts, etc). Instead, the procedure aims to extract data that is publicly available on the website. The data consists of the records of the following core variables. First, as the number of customers' questions might be regarded as a behavioral proxy of ``purchase intention'' (PI), we extracted each of these questions and counted them for a selected set of most commonly sold items across nations. Preliminary analyses revealed that four particular items emerged as the most frequent products traded in Mercadolibre (i.e., the Bible, Converse Chuck Taylor All-Stars, PlayStation 4, and Iphone7). Second, we also extracted the number of ``sold items'' (SI) as a proxy of sales. Third, we extracted the price of products as this is also an important element that drives sales. Fourth, we extracted the reputation of the vendor, as captured by the number of positive feedback (PF), neutral feedback (NeuF), and negative feedback (NF) that he or she has accumulated historically. Fifth, the trust index for each vendor is just calculated as the following ratio $PF/(PF + NeuF + NF)$. Sixth, because the number of sold items tends to increase as time goes by (i.e., sellers with more years of experience tend to have more sales than sellers with fewer years of experience) we also extracted the history of each seller that is expressed in years and months and analyzed it in terms of months. The pre-processing of data consisted of removing all non-numeric characters in the records of our core variables.

Our web scraping script allowed us to retrieve a total of 51,556 original records from all Latin-American countries with commercial operations in Mercadolibre. After pre-processing data, we discarded the records with missing values in the column of sold items (i.e., those without registered sales), as well as those products that appeared as ``used products'', as they might introduce noise in our analyses due to other factors that are not present in brand new products (e.g., poor quality or a significantly reduced price for its prolonged use before selling). After this preliminary analysis, we identified a total of 9,707 valid records that allowed us to described the statistical behavior of the trust scores for each country (see Figure 1A). Then, we selected the countries that showed a consistent statistical behavior in their trust scores (i.e., countries with highest trust scores with low statistical deviation). Following this criterion the resulting sample consisted of 8,292 records from Argentina, Brasil, Chile, Colombia, Ecuador, Mexico, Uruguay, and Venezuela (see Figure 1B).
We performed all these procedures as an easy-to-follow guide that interested readers can find and download in the following online repository \cite{Correa2021}. 

\subsection{Data Analysis}
We conducted our analyses in the R system \cite{RCore2019}, by employing data visualization techniques \cite{Chang2012}, as well as bivariate analysis and regression analysis for count data \cite{Cameron2013, Zeileis2008}. Regression analysis for count data is a non-linear regression model that, although is not well-known as other types of regression models, is convenient for analyzing non-negative integer-valued random variables assumed to be independently identically distributed (iid). According to \citeauthor{Cameron2013} \citeyear{Cameron2013} ``regression analysis of counts is motivated by the observation that in many, if not most, real-life contexts, the iid assumption is too strong'' (p. 2), so we may think of this technique as one that applies to the special case when the dependent variable, $y$, is restricted to be a non-negative random variable whose conditional mean depends on some vector of regressors, $x$. Poisson regression is the most commonly used technique for modeling this type of models. According to these considerations, the mathematical specification of our model shows the following form
\begin{equation}
\label{eq1}
    ln(\mu_{j})=\beta_{0}+\beta_{1}trust_{j}+\beta_{2}trust_{j}^{2}+\beta_{3}history_{j}+\beta_{4}ln(price_{j})+\beta_{5}PI_{j} + \delta
\end{equation}
where $ln(\mu_{j})$ is the natural logarithm of the mean or expected number of sales for the individual $j$ for $j = 1, ..., N$, and $\delta$ is the residual term. We estimate the parameters of our model though the use of maximum likelihood (ML) under the following equation   
\begin{equation}
    \label{eq2}
    P(SI_{j}=si_{j})=\frac{exp(-\mu_{j})\mu_{j}^{si_{j}}}{si_{j}!}
\end{equation}
Equation \ref{eq2} shows that the probability of $si_{j}$, the sale observed for individual $j$, follows a Poisson distribution for the mean of the counts of equation \ref{eq1}. A limitation of the Poisson distribution is overdispersion, and its presence makes it implausible to assume this type of distribution for the errors. In such circumstances, a negative binomial regression model is a viable alternative. We verify these assumptions through the likelihood ratio test. In any case, the emphasis is put on the statistical consistency of relations across countries. In this context, a relationship between any pair of variables is consistent if it shows the same sign (either positive or negative) and a reasonable magnitude that falls inside confidence intervals.

\section{Results}

\begin{figure}[h!]
    \centering
\includegraphics[height=.35\textwidth]{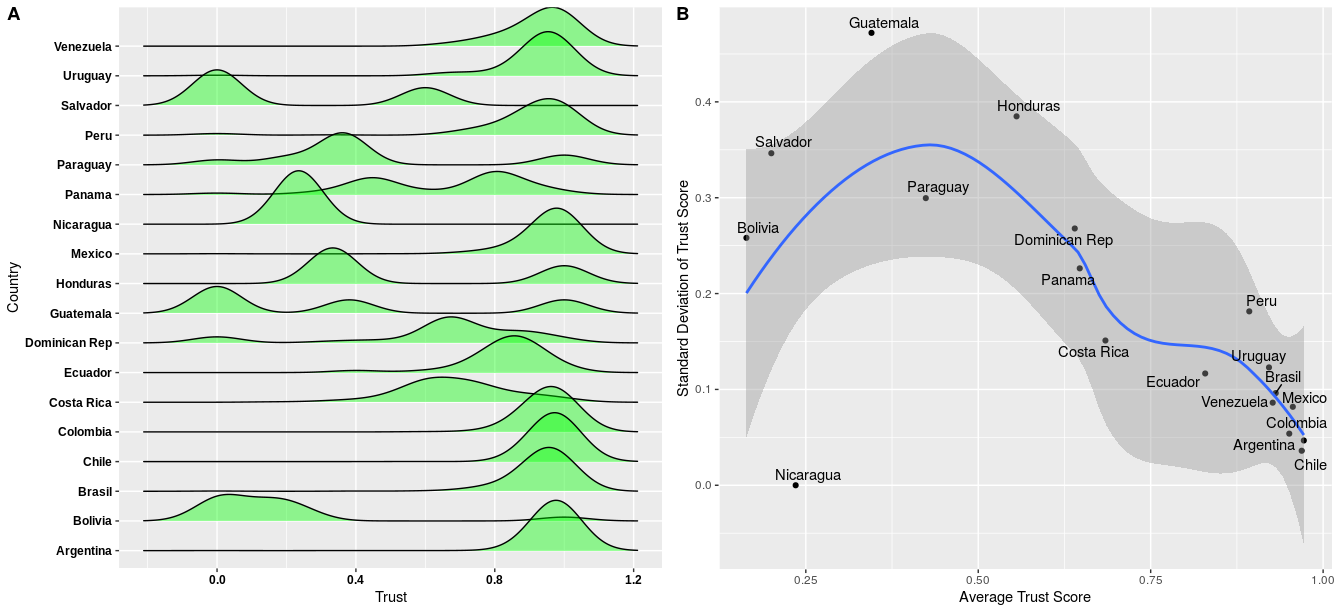}
    \caption{(A) Statistical distribution of trust scores for Mercadolibre vendors. (B) Relationship between the average and the standard deviation of trust scores in Mercadolibre}
    \label{FIG1}
\end{figure}

Figure \ref{FIG1}A shows the statistical behavior of the trust index. Among all nations, El Salvador, Bolivia, Nicaragua, Guatemala, Honduras, Paraguay, Dominican Republic, Panama, and Costa Rica revealed the lowest values (trust $\leq$ 0.70) with highly dispersed distributions (0.006 $\geq$ SD $\leq$ 0.436). In contrast, Argentina, Brasil, Chile, Colombia, Ecuador, Mexico, Uruguay, and Venezuela showed homogeneous distributions (0.046 $\geq$ SD $\leq$ 0.181) with the highest values (trust $\geq$ 0.85). We used this group of nations as a valid sample that allows us to examine the relationship between trust scores and sales (n = 8,292). Figure \ref{FIG1}B reveals that the average and the standard deviation of trust scores maintain a curve-like decreasing relationship with a significant non-parametric Spearman correlation ($\rho$ = -0.682, p $ = 0.0024$), meaning that, in general, countries with higher trust scores are those with less variance, excepting Nicaragua.

\begin{figure}[h!]
    \centering
\includegraphics[width=1\textwidth]{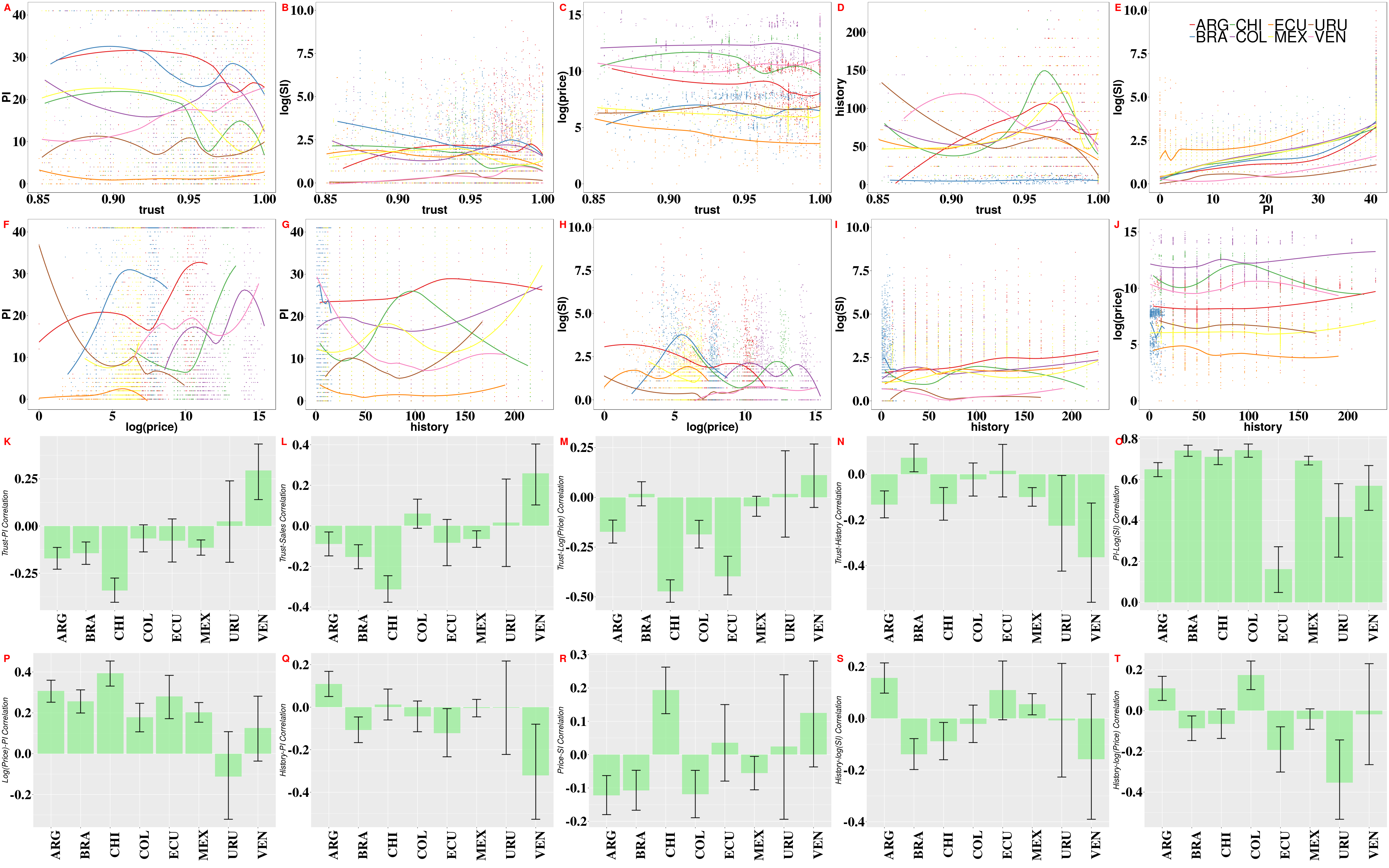}
    \caption{(A - J) Scatter plots show the relationship between pairs of core variables. Colors of fitted lines represent the relationship for each country. (A) Trust  and Purchase Intention (PI), (B) Trust and Sales, (C) Trust and price, (D) Price and sellers' history, (E) PI and sales, (F), Price and PI, (G) Sellers' history and PI, (H) Price and sales, (I) Sellers' history and sales, (J) Sellers' history and price. (K - T) bar plots with confidence intervals of correlations between pairs of variables. (K) Magnitude of the correlation between trust and PI. (L) Magnitude of the correlation between trust and sales, (M) Magnitude of the correlation between trust and price, (N) Magnitude of the correlation between trust and history, (O) Magnitude of the correlation between PI and sales, (P) Magnitude of the correlation between price and PI, (Q) Magnitude of the correlation between history and PI, (R) Magnitude of the correlation between price and sales, (S) Magnitude of the correlation between history and sales, (T) Magnitude of the correlation between history and price.}
    \label{FIG2}
\end{figure}

Figure \ref{FIG2} provides a useful visual exploration of bivariate relationships.

This visual exploration focuses on the relation that sales (SI) have with four of its data-driven predictors: purchase intention (PI), price, trust, and sellers' history. Given the large statistical variance of both SI and price, we applied a log-transformation to their raw values for the sake of data visualization and parameters estimation. Figure \ref{FIG2} shows these sets of relationships with colored scatter plots that differentiate their estimated non-linear correlations across countries, as well as the magnitude of these correlations and their corresponding 95\% confidence intervals for each country. None of the correlations were statistically consistent among all nations analyzed, despite being statistically significant. There was, however, an interesting exception. The relationship between purchase intention (PI) and sales (SI), illustrated in Figure \ref{FIG2}E and \ref{FIG2}O, was the only one that showed a consistent pattern regardless of the country (i.e., the correlation was always positive and statistically significant). To dive deeper into these results, the attention goes to a series of regression results. Table \ref{T2} shows the results from six Poisson regression model.

\begin{table}[!htbp] \centering 
  \caption{Poisson Regression Model} 
  \label{T2} 
   \begin{tabular}{lcccccc}
\hline
\multicolumn{7}{c}{Dependent variable: SI}                                                                                                                                                                         \\ \cline{2-7} 
                       & (1)                          & (2)                           & (3)                           & (4)                          & (5)                          & (6)                          \\
\textit{trust}                  & 8.100***                     & 17.369***                     & 18.330***                     & -0.085***                    & 5.668***                     & 257.780***                   \\
                       & (0.119)                      & (0.250)                       & (0.314)                       & (0.026)                      & (0.226)                      & (10.230)                     \\
\textit{trustsq}          & -5.149***                    & -10.527***                    & -10.625***                    &                              & -3.568***                    & -162.267***                  \\
                       & (0.075)                      & (0.148)                       & (0.181)                       &                              & (0.134)                      & (6.086)                      \\
\textit{history}                &                              & 0.001***                      & 0.002***                      & 0.004***                     & 0.004***                     & 0.195***                     \\
                       &                              & (0.000)                       & (0.000)                       & (0.000)                      & (0.000)                      & (0.001)                      \\
\textit{log(price)}                &                              &                               & -0.128***                     & -0.267***                    & -0.268***                    & -12.167***                   \\
                       &                              &                               & (0.001)                       & (0.000)                      & (0.001)                      & (0.044)                      \\
\textit{PI}                     &                              &                               &                               & 0.107***                     & 0.107***                     & 4.879***                     \\
                       &                              &                               &                               & (0.000)                      & (0.000)                      & (0.013)                           \\
\textit{Constant}               & \multicolumn{1}{c}{0.736***} & \multicolumn{1}{c}{-3.244***} & \multicolumn{1}{c}{-3.115***} & \multicolumn{1}{c}{2.515***} & \multicolumn{1}{c}{0.287***} & \multicolumn{1}{c}{} \\
                       & \multicolumn{1}{c}{(0.047)}  & \multicolumn{1}{c}{(0.105)}   & \multicolumn{1}{c}{(0.136)}   & \multicolumn{1}{c}{(0.025)}  & \multicolumn{1}{c}{(0.096)}  & \multicolumn{1}{c}{}  \\
Observations           & \multicolumn{1}{c}{7682}     & \multicolumn{1}{c}{7327}      & \multicolumn{1}{c}{6219}      & \multicolumn{1}{c}{6219}     & \multicolumn{1}{c}{6219}     & \multicolumn{1}{c}{6219}     \\
Prob \textgreater $\chi^2$ & \multicolumn{1}{c}{0.0000}   & \multicolumn{1}{c}{0.0000}    & \multicolumn{1}{c}{0.0000}    & \multicolumn{1}{c}{0.0000}   & \multicolumn{1}{c}{0.0000}   & \multicolumn{1}{l}{}         \\
Pseudo $R^{2}$              & \multicolumn{1}{c}{0.0054}   & \multicolumn{1}{c}{0.0058}    & \multicolumn{1}{c}{0.0253}    & \multicolumn{1}{c}{0.4070}   & \multicolumn{1}{c}{0.4077}   & \multicolumn{1}{l}{}         \\
AIC.      & \multicolumn{1}{c}{1646944}  & \multicolumn{1}{c}{1615727}   & \multicolumn{1}{c}{1442599}   & \multicolumn{1}{c}{877707.4} & \multicolumn{1}{c}{876634.1} & \multicolumn{1}{l}{}         \\ \hline
\multicolumn{6}{l}{\begin{tabular}[c]{@{}l@{}}
\end{tabular}} \\
\multicolumn{7}{l}{\begin{tabular}[c]{@{}l@{}}Note: Standard errors between parentheses; Delta-method in column 6 (Marginal effects).\\ Statistical significance:*p\textless{}0.1, **p \textless{}0.05, ***p \textless{}0.01. \end{tabular}}
\end{tabular}
\end{table}

\begin{figure}[h!]
    \centering
\includegraphics[width=0.4\textwidth]{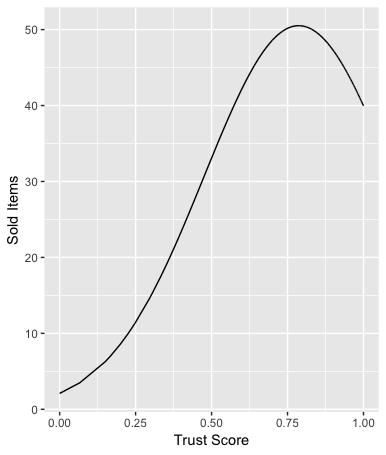}
\caption{This figure plots equation 1 to show the relationship between SI and trust.}
    \label{FIG3}
\end{figure}

Here, we find the existence of a quadratic-like relationship between sales (SI) and \textit{trust}. While the sign of the coefficient of \textit{trust} is positive, the sign of the coefficient of trust square (\textit{trustsq}) is negative. To interpret these results, suppose we use the coefficients from regression \ref{eq1} to plot the following fitted equation $SI = \exp(0.736 + 8.100*trust - 5.149*trust^{2})$. Figure \ref{FIG3} depicts the relationship between the empirical Trust index and sales, as described by the fitted equation. Here we show that sales, as captured by the number of sold items (SI), increase with higher values of \textit{trust}, but such an increment reaches a maximum when \textit{trust} is around 0.80. After this point, sales decrease. In regressions (2)-(5), we gradually include different control variables. The signs of the coefficients for the control variables \textit{history}, \textit{price}, and purchase intention (\textit{PI}) were as expected. Column (6) shows the average marginal effects. \textit{trust} was associated with a bigger chance in sales than any other variable and it was robust to any specification, always showed a substantial impact on sales and was highly significant. However, this relationship is significant only when the model includes its quadratic form; \textit{trust} is not consistent otherwise (see column 4).

To verify the statistical assumption that the distribution of errors is a Poisson one, we estimate a model with pseudo-ML.

\begin{table}[h!]
\centering 
  \caption{Negative Binomial Regression} 
  \label{T3} 
\begin{tabular}{lcccccc}
\hline
\multicolumn{7}{c}{Dependent variable: SI}                                                           \\ \cline{2-7} 
                           & (1)       & (2)       & (3)       & (4)       & (5)       & (6)         \\ \cline{2-7} 
\textit{trust}                      & 5.743***  & 6.733***  & 6.830***  & -4.362*** & 10.012*** & 432.772***  \\
                           & (0.578)   & (0.894)   & (1.030)   & (0.264)   & (0.860)   & (41.638)    \\
\textit{trustsq}              & -3.761*** & -4.326*** & -4.120*** &           & -9.240*** & -399.436*** \\
                           & (0.458)   & (0.633)   & (0.747)   &           & (0.595)   & (30.972)    \\
\textit{history}                    &           & 0.001***  & 0.002***  & 0.003***  & 0.003***  & 0.135***    \\
                           &           & (0.000)   & (0.000)   & (0.000)   & (0.000)   & (0.014)     \\
\textit{log(price)}                    &           &           & -0.114*** & -0.168*** & -0.163*** & -7.058***   \\
                           &           &           & (0.008)   & (0.006)   & (0.006)   & (0.350)     \\
\textit{PI}                         &           &           &           & 0.092***  & 0.093***  & 4.042***    \\
                           &           &           &           & (0.001)   & (0.001)   & (0.149)     \\
\textit{Constant}                   & 1.711***  & 1.230***  & 1.761***  & 6.331***  & 0.997***  &             \\
                           & (0.180)   & (0.322)   & (0.362)   & (0.243)   & ()        &             \\
Observations               & 7682      & 7327      & 6219      & 6219      & 6219      &             \\
Prob \textgreater chi2     & 0.0000    & 0.0000    & 0.0000    & 0.0000    & 0.0000    &             \\
Pseudo R2                  & 0.0013    & 0.0008    & 0.0046    & 0.1167    & 0.1202    &             \\
AIC.          & 62386.41  & 59825.14  & 50671.51  & 44969.04  & 44790.2   &             \\
LR test \\
$\alpha$ (p-value) & 0.000     & 0.000     & 0.000     & 0.000     & 0.000     &             \\ \hline
\multicolumn{7}{l}{\begin{tabular}[c]{@{}l@{}}Note: Standard errors between parentheses; Delta-method in column 6 (Marginal effects).\\ Statistical significance:*p\textless{}0.1, **p \textless{}0.05, ***p \textless{}0.01. \end{tabular}}
\end{tabular}
\end{table}
\newpage
The results show that the standard errors are much smaller, pointing out that they do not follow a Poisson distribution. This result leads us to employ a negative binomial (NB) regression model to obtain more consistent results (see Table \ref{T3}). The likelihood ratio test of alpha indicates that there are problems of overdispersion in the models, confirming the use of a Negative Binomial regression model as the most convenient specification for the case at hand. After correcting the problem of errors, the results in Table \ref{T2} proved to be confirmed; namely, we found evidence pointing out the significant and robust association that trust shows with Latin-American vendors' sales in Argentina, Brasil, Chile, Colombia, Ecuador, Mexico, Uruguay, and Venezuela.

\section{Discussion}
A fundamental factor that drives customers to purchase products through the Internet is their trust in e-commerce platforms and their trust in vendors' reputations \cite{Hoffman1999,Cheung2000,Grabner2003}. Even though the relationship between trust and sales has been systematically analyzed in different countries \cite{Chevalier2006,cui2012}, we did not find previous similar efforts that analyzed this relationship sampling Latin American countries. 

Indeed, existent studies from Latin America have shown different orientations. In Chile, the study of \citeauthor{Bianchi2012} \citeyear{Bianchi2012} analyzed the influence of consumer’s perceptions of risk and trust on their attitudes and intentions to purchase on the Internet but did not cover sales. In Colombia, neither the study of \citeauthor{dakduk2017} \citeyear{dakduk2017} nor the work of \citeauthor{Sanchez2017} \citeyear{Sanchez2017} focused on sales. The comparison between Hispanics and Anglos by \citeauthor{changchit2009} \citeyear{changchit2009} did not focus on sales either. To the best of our knowledge, the most proximate study that focused on the trust-sales relationship was that of \citeauthor{LopezJimenez2021} \citeyear{LopezJimenez2021}. Still, its focus was on European and Latin American companies with commercial operations through the internet and adhered to an assurance seal as reliable and safe providers. Among their results, they found that the use of trust seals increased online sales for more than 66\%, improved their corporate image, enhanced the number of potential buyers who visit their websites, and generally captured consumers’ attention. 

The current study provided several contributions to the literature. First, it developed a conceptual and methodological data-driven framework to the trust-sales relationship. Rather than relying on traditional perspectives mostly inspired by dominant theories (TAM, TPB, UTAUT) and operationalized through surveys and questionnaires, the developed approach is admittedly more consistent with a business data analytic framework \cite{Provost2013}. Here, it is worth mentioning that the framework itself is not novel because previous efforts have also harnessed this data-driven orientation \cite{correa2019, Teichert2020}. Nonetheless, the originality of the current work relies on its object of study and its emphasis on evaluating the statistical consistency of the trust-sale relationship. Such an emphasis has been neglected in the literature of Latin American e-commerce. The rigor present in the statistical analyses through data visualization techniques \cite{Chang2012} combined with regression models for count data \cite{Cameron2013,Zeileis2008} and corresponding supplemental material also offers valuable tools for interested researchers. These tools aim to replicate the findings here reported and help other understand additional computational details that go beyond the scope of this article.

The results of the current work revealed unprecedented findings in the literature.  First, the study showed two groups of Latin-American nations in terms of the trust index analyzed. On the one hand, Argentina, Brasil, Chile, Colombia, Ecuador, Mexico, Uruguay, and Venezuela emerged as a cluster of nations that revealed the most homogeneous distributions with the highest trust indexes in the region. On the other hand, El Salvador, Bolivia, Nicaragua, Guatemala, Honduras, Paraguay, Dominican Republic, Panama, and Costa Rica revealed the lowest trust indexes with highly dispersed distributions. The existence of these clusters suggests a differentiated adoption of e-commerce in Latin America. Thus, customers' trust in local vendors seems to be associated with nationwide effects. The relevance of this finding should be clear for further research that targets an empirical sample in one country only as compared with those targeting multi-country samples.

The current work extended some lessons from recent studies focusing on boosting sales and customers' satisfaction with e-commerce \cite{Lin2019}. The fact that trust proved to be associated with a bigger chance in sales than any other variable, revealing a quadratic-like form relationship  posits interesting practical implications. At least in Latin America, online sales are neither increased by a lengthy experience history of a seller nor with competitive cheaper prices. What seems to be critical from the customer's point of view is to transact with a reliable seller whose feedback shows how other customers were attended before and after the commercial transaction. As the trust score in Mercadolibre is intended to capture the reliability of a seller, the closer the seller's trust score to 100\%, the higher the probability of increasing sales in Mercadolibre. In other words, the main sign of confidence comes from the informational balance made by the consumer when he or she reads the comments or feedback of other customers before deciding to buy an item from a seller. Here, two mechanisms might be differentiated. On the one hand, the number of questions that other buyers post to the seller before initializing the transaction (i.e., our proxy to purchase intention) might serve as an essential but subtle mechanism that drives the purchase decision in Mercadolibre. On the other hand, the feedback that other customers have given to the seller after completing the transaction defines the reliable conduct of a seller on the website. This feedback, seen as a customer's satisfaction indicator, is a powerful social influence that frames all other aspects of the website, because feedback conveys information about customers' perceived value, expectations, perceived quality and buyer loyalty to varying degrees, which closely mirrors what happens in offline shopping environments \cite{Nisar2017}. In this sense, our results are aligned with those reported by \citeauthor{Wang2017} \citeyear{Wang2017} who also observed the influence of feedback and word-of-mouth on consumers' intention to buy a product and sharing product information with others on social commerce websites. 

Finally, but not least importantly, our work illustrated the heuristic value of Mercadolibre's current practices to customers' trust by depicting a sensitive index whose meaning is easy to understand for all customers. We regard that this heuristic value might inspire the academic community to rethink the conceptualization and operationalization of fundamental concepts beyond traditional frameworks that have been recognized as confusing or ill-defined \cite{Grabner2003,Connolly2007,Connolly2008}. 

The current work is not free of limitations. Given the geographically-oriented focus of this work, the consistency of trust-sales relationship in other regions remains as an open question for further research. We foresee that in the near future other works will generalize our procedure to answer this question by harnessing web scraping procedures to extract data from Amazon (Global), Tmall Global (Asia), Shopify (Europe), Jumia (Africa) or Woolworths (Oceania). The exploration of the variables available by each of these platforms is another interesting question that deserves an empirical analysis. A comparison of similar trust indexes might be revealing in terms of sellers’ conduct differences across regions. While the present study did not explore these questions, further endeavors can use this work for theory-building purposes. As an associated orientation, we highlight the use of the trust index as a valid criterion for further endeavors to develop new surveys or questionnaires of customers’ trust.

\end{document}